\documentclass{elsart}
\usepackage{graphicx}
\usepackage{subfigure}

\def\ba{\begin{eqnarray}}
\def\ea{\end{eqnarray}}

\begin{document}

\begin{frontmatter}

\title{\bf Hydrodynamic predictions for Pb+Pb collisions at $\sqrt{s_{\rm NN}}$ = 2.76 TeV
} 

\author[rzu,ifj]{Piotr Bo\.zek},
\author[ifj]{Mikolaj Chojnacki},
\author[ujk,ifj]{Wojciech Florkowski}, and
\author[bb,ctu]{Boris Tom\'a\v{s}ik} 

\address[rzu]{Institute of Physics, Rzesz\'ow University, ul. Rejtana 16, 
35-959 Rzesz\'ow, Poland}
\address[ifj]{The H. Niewodnicza\'nski Institute of Nuclear Physics, Polish Academy of Sciences, ul. Radzikowskiego 152, 31-342 Krak\'ow, Poland}
\address[ujk]{Institute of Physics, Jan Kochanowski University,
ul.~\'Swi\c{e}tokrzyska 15, 25-406~Kielce, Poland} \address[bb]{Univerzita Mateja Bela, Tajovsk\'eho 40, 97401 Bansk\'a Bystrica, Slovakia}
\address[ctu]{FNSPE, Czech Technical University in Prague, B\v{r}ehov\'a 7, 11519 Prague, 
Czech Republic}

\begin{abstract} 
Using the newest data for $pp$ scattering at the CERN Large Hadron Collider  
(LHC) combined with the Glauber model, we make hydrodynamic predictions for the soft hadronic observables planned to be measured in the forthcoming Pb+Pb collisions at $\sqrt{s_{\rm NN}}$ = 2.76 TeV.
\end{abstract}
\end{frontmatter}
\vspace{-7mm} PACS: 25.75.-q, 25.75.Dw, 21.65.Qr

{\bf 1.} The behavior of matter created in  relativistic heavy-ion collisions at the BNL Relativistic Heavy Ion Collider (RHIC) is very well described by the perfect-fluid hydrodynamics \cite{Kolb:2003dz,Huovinen:2003fa,Shuryak:2004cy,Teaney:2001av,Hama:2005dz,Hirano:2007xd,Nonaka:2006yn,Chojnacki:2007rq,Bozek:2009ty}. Having in mind these successes, it is appealing to use perfect-fluid hydrodynamics to make predictions for the Pb+Pb collisions at the LHC. Several such studies have been performed so far and they are summarized in Ref. \cite{Abreu:2007kv}. 

Nevertheless, in the last two years  two important aspects concerning the heavy-ion physics at the LHC have  changed: Firstly, the forthcoming heavy-ion experiments will have a lower energy than that planned previously,  \mbox{$\sqrt{s_{\rm NN}}$ = 2.76} TeV instead  of \mbox{$\sqrt{s_{\rm NN}}$ = 5.5 TeV}. Secondly, the data for $pp$ scattering at the LHC have just been  analyzed. These  new facts suggest that a revised  prediction for \mbox{$\sqrt{s_{\rm NN}}$ = 2.76 TeV}  based on the newest $pp$ results is advisable. 

In the paper we follow this idea and use the newest data for $pp$ scattering at the LHC to make hydrodynamic predictions for  soft hadronic observables measured in  Pb+Pb collisions at $\sqrt{s_{\rm NN}}$ = 2.76 TeV. The $pp$ data are rescaled by the number of sources in heavy-ion collisions (estimated from the Glauber model) in order to determine the final multiplicity of charged hadrons in nuclear collisions. The relative increase of the  multiplicity of particles  produced at RHIC and LHC energies requires a similar increase of the initial entropies. This, in turn, allows us to make estimates of the initial central temperature expected at the LHC.

Our hydrodynamic approach is defined in Ref. \cite{Chojnacki:2007rq}, where we have  described quite successfully the RHIC data and made predictions for Pb+Pb collisions at $\sqrt{s_{\rm NN}}$ = 5.5 TeV. It is a 2+1 dimensional (boost-invariant) hydrodynamic model with a  realistic equation of state based on the lattice results and the hadron-gas model. For simplicity, a single-freeze-out scenario  is adopted in the Monte-Carlo version \cite{Kisiel:2005hn} --- all known hadrons are generated on the freeze-out hypersurface determined by the condition of constant freeze-out temperature, then, unstable resonances decay with the branching ratios given by the PDG. The hadronic interactions in the final state are neglected. 

{\bf 2.} In order to specify the initial entropy density for our hydrodynamic calculations we determine first the expected multiplicity in nuclear collisions. In Ref. \cite{McLerran:2010ex} the  data on $dN^{pp}/d\eta |_{\eta=0}$ from the first proton-proton collisions at the LHC  \cite{:2009dt,Aamodt:2010ft,Aamodt:2010pp,Khachatryan:2010xs,Khachatryan:2010us}
have been fitted with the parameterization 
\begin{equation}
\left. \frac{dN^{pp}(E)}{d\eta} \right|_{\eta=0}  = 3.70347 \, 
\left ( \frac{E}{1\, \mbox{TeV}}\right )^{0.23},
\label{larry}
\end{equation}
where $E$ is the CMS energy of the collision. For $E$ = 2.76 TeV this fit gives $dN^{pp}/d\eta |_{\eta=0} = 4.68$. We shall, however, rather base our estimate on the closest data point which has been measured by ALICE \cite{Aamodt:2010ft} and the CMS  Collaboration \cite{Khachatryan:2010us} at $E= 2.36$~TeV. The weighted average of the two results with weights given by the inverse of statistical uncertainties gives $dN^{pp}(2.36)/d\eta |_{\eta=0} = 4.65$. This result is extrapolated with the power-law prescription (\ref{larry}) to 
\begin{equation}
\left. \frac{dN^{pp}(2.76)}{d\eta} \right|_{\eta=0}  = 
\left ( \frac{2.76}{2.36}\right )^{0.23} 4.65 = 4.82.
\label{multPP276}
\end{equation}
This value will be used in our further calculations. 

The expected multiplicity in Pb+Pb collisions is obtained by the multiplication of the $pp$ result by the number of sources that are identified with the wounded (participant) nucleons and binary collisions
\begin{eqnarray}
\left. \frac{dN^{PbPb}(b,2.76)}{d\eta} \right|_{\eta=0} = 
\left. \frac{dN^{pp}(2.76)}{d\eta} \right|_{\eta=0} \, N_{\rm sr}(b) 
= 4.82 \, N_{\rm sr}(b)\, ,
\label{scaling}
\end{eqnarray}
\begin{eqnarray}
N_{\rm sr} = \frac{1-\alpha}{2}\, N_{\rm w} + \alpha \,N_{\rm bin}\, .
\label{sources}
\end{eqnarray}
Here the number of participant nucleons $N_{\rm w}$,  binary collisions $N_{\rm bin}$, and the number of sources $N_{\rm sr}$ depend on centrality and  are obtained  from the GLISSANDO code \cite{Broniowski:2007nz} that is a Monte-Carlo implementation of the Glauber model. The parameter $\alpha$ in (\ref{sources}) takes into account the contribution of minijets to the fireball density. The jet contribution is expected to rise  with the energy of the collisions.  The centrality studies performed by PHOBOS on Au+Au collisions gave  $\alpha=0.12$ at $\sqrt{s_{NN}} = 19.6$~GeV  and  $\alpha=0.145$ at  $200$~GeV \cite{Back:2004dy}. In the present analysis, assuming the linear growth of $\alpha$ with $\log(s)$, we take $
\alpha = 0.16$.

\begin{figure}[b]
\begin{center}
\includegraphics[width=.7\textwidth]{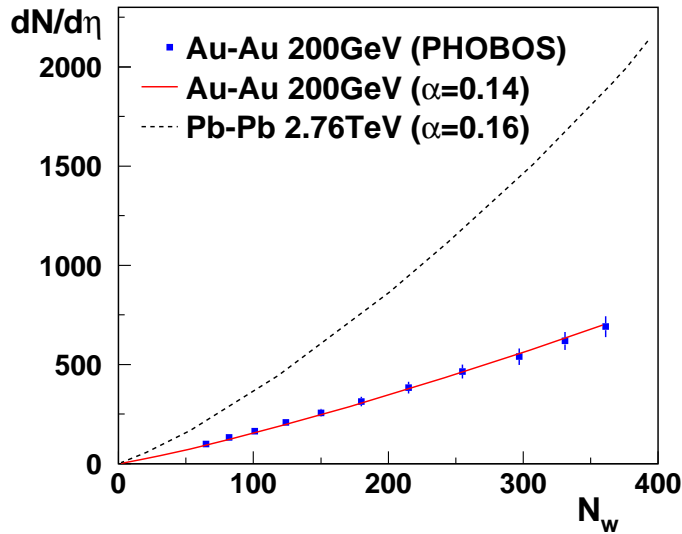}
\caption{Charged particle multiplicity density in pseudorapidity for Au+Au collisions at $\sqrt{s_{\rm NN}}=200$ GeV (solid line) together with the PHOBOS data \cite{Back:2001xy}, and predictions for Pb+Pb collisions at $2.76$ TeV using $\alpha=0.16$. The model calculations are performed with GLISSANDO \cite{Broniowski:2007nz} and use Eqs.~(\ref{scaling}) and (\ref{sources}). }
\label{fig:dndeta}
\end{center}
\end{figure}
The centrality dependence of the charged particle pseudorapidity density  follows from the centrality dependence of $N_{\rm sr}(b)$ that is obtained from GLISSANDO with $NN$ cross section of 63~mb. It is presented in Fig. \ref{fig:dndeta}.

{\bf 3.} For our hydrodynamic simulations we need the initial profiles of the entropy density. Following the standard approach used in hydrodynamic models we assume that the initial entropy density is proportional to the source density in the transverse plane
\begin{equation} 
s(b,x_\perp) = s(T_i) \, \frac{\rho_{\rm sr} (b,x_\perp)}{\rho_{\rm sr} (0,0)}.
\label{entropy1}
\end{equation}
where $s(T_i)$ is the central entropy density at zero impact parameter. The integral of the source density defines $N_{\rm sr}$,
\begin{equation}
N_{\rm sr}(b) = \int d^{\,2} x_\perp \, \rho_{\rm sr} (b,x_\perp) \, .
\label{source_density}
\end{equation}
Unfortunately, GLISSANDO does not provide the source densities which are sufficiently smooth and could be directly used as initial conditions for our simulations. We thus calculate them in the optical Glauber approach in which we fix the impact parameter $b$ in such a way that we reproduce the number of sources obtained in GLISSANDO, in agreement with Eq. (\ref{source_density}). 

The absolute scale of the entropy density is estimated from the ratio of the multiplicity densities. We assume that the ratios of  total entropies in the transverse plane (at midrapidity and at the same initial proper time) should be equal to the ratios of the multiplicities. By considering the ratios of the entropies for the LHC (at $\sqrt{s_{\rm NN}}$ = 2.76 TeV) and RHIC (at $\sqrt{s_{\rm NN}}$ = 200 GeV) we find
\begin{equation}
\!\!\!\!
\frac{S_{PbPb}(b,2.76)}{S_{AuAu}(b,0.200)} 
=
\frac{s(T_i^{PbPb}) N^{PbPb}_{\rm sr}(b) \rho^{AuAu}_{\rm sr} (0,0) }
{s(T_i^{AuAu})  N^{AuAu}_{\rm sr}(b) \rho^{PbPb}_{\rm sr} (0,0)}
=
\frac{\left. \frac{dN^{PbPb}(b,2.76)}{d\eta} \right|_{\eta=0}}
{\left. \frac{dN^{AuAu}(b,0.200)}{d\eta} \right|_{\eta=0}} .
\label{ratio1}
\end{equation}
Expressing the multiplicity density at the LHC with the help of Eq.~(\ref{scaling})
we obtain the ratio of the entropy densities
\begin{eqnarray}
\frac{s(T_i^{PbPb})}{s(T_i^{AuAu})} 
= 
\frac{ \rho^{PbPb}_{\rm sr} (0,0)}
{\rho^{AuAu}_{\rm sr} (0,0)} \,\,
\frac{ 4.82  \,\, N^{AuAu}_{\rm sr}(b) }
{  \left. \frac{dN^{AuAu}(b,0.200)}{d\eta} \right|_{\eta=0}} .
\label{ratio2}
\end{eqnarray}
The calculations of the source densities give
\begin{eqnarray}
\rho^{PbPb}_{\rm sr} (0,0) = 6.76 \ \mbox{fm}^{-2} , \quad
\rho^{AuAu}_{\rm sr} (0,0) = 4.71 \ \mbox{fm}^{-2}.
\end{eqnarray}
For central Au+Au collisions at $\sqrt{s_{NN}}= 200$~GeV 
 (centrality class 0--5\%) we find $N^{AuAu}_{\rm sr}=308$ and the
 PHOBOS multiplicity density  is 667 \cite{Back:2001xy}. The ratio of
 entropies in Eq. (\ref{ratio1}) is then 3.2, and the ratio in
 Eq.~(\ref{ratio2}) acquires practically the same 
value~\footnote{For larger values of $b$, using the Glauber 
estimates for $N^{AuAu}_{\rm sr}$ and the experimental values for
 $dN^{AuAu}/d\eta \, |_{\eta=0}$, we find deviations from 3.2 by 
about 10\%. Such differences indicate that the scalings (\ref{larry})
 and (\ref{entropy1}) are valid with a similar accuracy.}. The transverse size of the fireball $S_\perp=\pi \sqrt{\langle x^2\rangle \langle y^2\rangle}=22.9$~fm$^2$
 for the most central Au+Au 
collisions at RHIC  is almost the
 same as for central Pb+Pb collisions  at the LHC: $S_\perp=23.2$~fm$^2$.

The entropy density is connected to the temperature through the equation of state  constructed in \cite{Chojnacki:2007jc}. The multiplicity  in Au+Au collisions at \mbox{$\sqrt{s_{NN}}=200$} GeV is reproduced if the initial temperature at $\tau_i = 0.25$~fm is 520~MeV. From this we derive the initial temperature for $\sqrt{s_{NN}} = 2.76$~TeV to be 735~MeV at the initial proper time 
$\tau_i=0.25$~fm (the one used in the present paper), or 
$480$~MeV if $\tau_i=1$~fm.

\begin{figure}[t]
\begin{center}
\includegraphics[angle=0,width=0.7\textwidth]{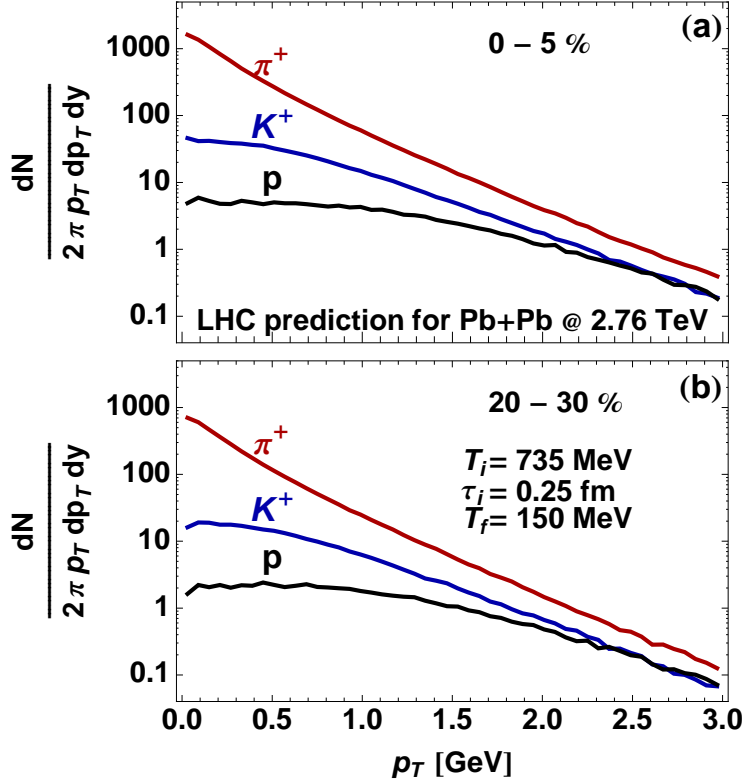}
\end{center}
\caption{\small Predictions for transverse-momentum spectra of pions, kaons, and protons for the two centralities: $c$=0--5\% (upper part) and $c$=20--30\% (lower part). The values of the input parameters are discussed in the text.}
\label{fig:sp}
\end{figure}

\begin{figure}[t]
\begin{center}
\includegraphics[angle=0,width=0.7\textwidth]{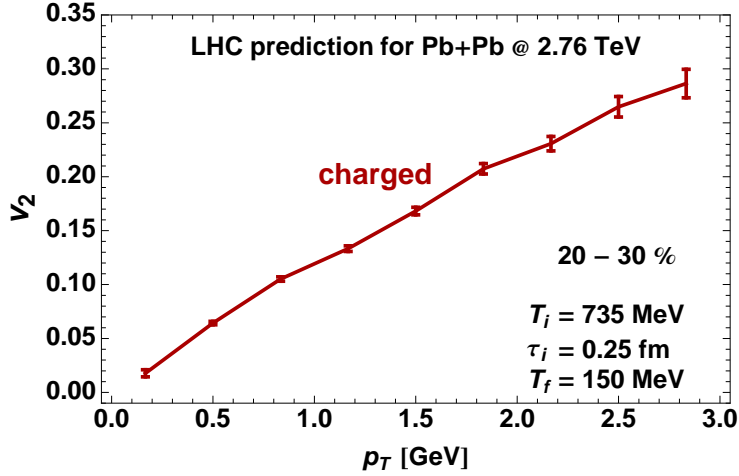}
\end{center}
\caption{\small The elliptic flow coefficient $v_2$ of charged 
particles shown as a function of the transverse momentum. 
The parameters the same as in the lower part of Fig. \ref{fig:sp}. }
\label{fig:v2}
\end{figure}

\begin{figure}[thb]
\begin{center}
\includegraphics[angle=0,width=0.7\textwidth]{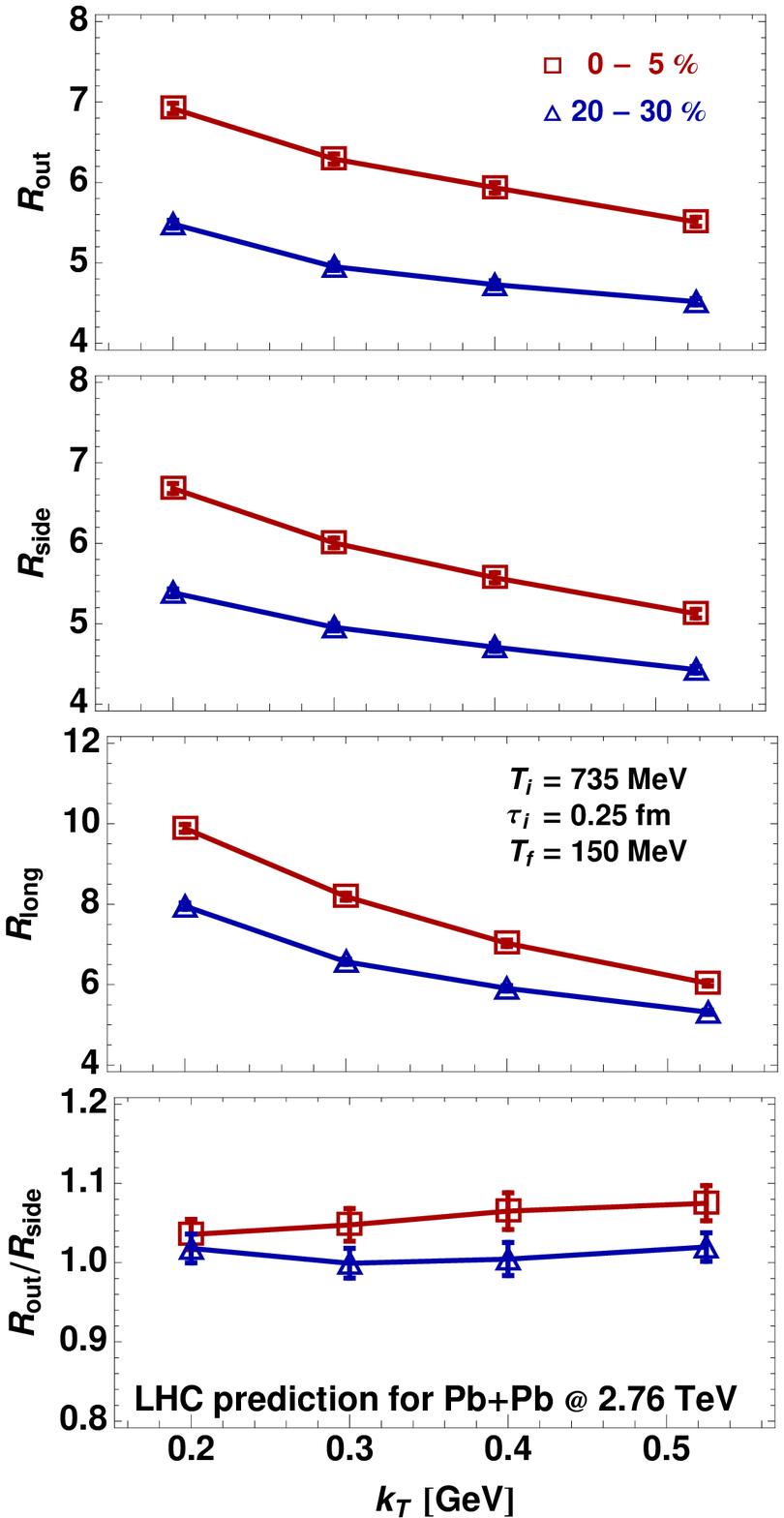}
\end{center}
\caption{\small The pionic HBT radii shown as functions of the transverse momentum of the pairs for the central (squares) and non-central (triangles) collisions.}
\label{fig:hbt}
\end{figure}

{\bf 4.} In Figs. \ref{fig:sp}--\ref{fig:hbt} we present our results for: the transverse-momentum spectra of pions, kaons, and protons, the elliptic flow of charged particles, and the HBT radii of pions. The calculations are done for two centrality classes (\mbox{c = 0--5\%} and c = 20--30\%).  The hydrodynamic calculations use the following parameters: the starting time of the hydrodynamic evolution \mbox{$\tau_i$ = 0.25 fm}, the initial central temperature (in the most central collisions) \mbox{$T_i =$  735 MeV}, and the freeze-out temperature \mbox{$T_f =$ 150} MeV. From the measured $\bar{p}/p$ ratios in proton-proton
collisions at the LHC 
\cite{Aamodt:2010dx} we estimate that $\bar{p}/p=0.97$ in nuclear collisions at $\sqrt{s_{NN}}=2.76$ TeV, which corresponds to a small value of the baryon chemical potential $\mu_B \simeq 2.2$~MeV, if no additional baryon stopping occures in 
Pb+Pb collisions. In our calculations all chemical potentials in THERMINTOR have been 
set equal to zero.

The spectra shown in Fig. \ref{fig:sp} yield the following mean values of the transverse momentum: $\langle p^{\rm pions}_\perp \rangle$ = 656 MeV, 
$\langle p^{\rm kaons}_\perp \rangle$ = 1028 MeV, and \mbox{$\langle p^{\rm protons}_\perp \rangle$ = 1460} MeV  for the centrality class $c$ = 0--5\%\,,  and $\langle p^{\rm pions}_\perp \rangle$ = 637 MeV, \mbox{$\langle p^{\rm kaons}_\perp \rangle$ = 997} MeV, and \mbox{$\langle p^{\rm protons}_\perp \rangle$ = 1406} MeV for the centrality class $c$ = 20--30\%. 
The mean transverse momentum increases from 
the top RHIC energies to the LHC energies by  about $35-40$\%.
 The expected pseudorapidity (rapidity) density of 
charged particles is $dN/d\eta = 2161$ ($dN/dy = 2457$) for  $c$ = 0--5\%, and 905 (1036) for $c$ = 20--30\%. More specifically we predict
\begin{equation}
\frac{dN}{dy}= 930, \ 146, \ \ \mbox{and}\ \ 69, 
\end{equation}
for $\pi^+$, $K^+$ and protons respectively, for central collisions.
The rise of  the entropy from RHIC to LHC and the similarity of the 
transverse sizes of the fireballs leads to an increase of the upper range 
of the scaling density \cite{Voloshin:2007af}
from the value  
\begin{equation}
\frac{1}{S_\perp} \frac{dN}{dy}\simeq 35\  \mbox{fm}^{-2} 
\end{equation}
at RHIC energies, to 
\begin{equation}
\frac{1}{S_\perp} \frac{dN}{dy}\simeq 105\  \mbox{fm}^{-2} 
\end{equation}
for the central Pb+Pb collisions at $\sqrt{s_{NN}}=2.76$~GeV.

In Fig. \ref{fig:v2} the elliptic-flow coefficient $v_2$ 
of charged particles is shown as a function of transverse-momentum. 
Compared to the analogous RHIC plots, the dependence of $v_2$ on $p_\perp$
 exhibits 
saturation with increasing beam energy, as observed by Kestin and Heinz 
\cite{Kestin:2008bh}.  However, the substantial increase of the 
initial entropy  density (matching the expected rapid growth 
of the multiplicity with energy) and the hard equation of state used in the calculations lead to a significant increase of $\langle p_\perp \rangle$ and 
the integrated elliptic flow when going from from RHIC to the LHC. 
We find $v_2=0.088$ for charged particles in the 
centrality class $c=20-30$\%. The elliptic flow is known to be
very sensitive to shear viscosity effects \cite{Luzum:2009sb} that 
reduce its value. However, at the first stage of the expansion at the LHC, where stress
corrections are the most important, we deal with high temperature and  the 
relative importance of viscosity corrections is reduced compared to RHIC 
energies.

Figure \ref{fig:hbt} shows our predictions for the HBT radii. The radii are calculated with the methods defined in Ref. \cite{Kisiel:2006is}. Here again, a small increase of the radii, as compared to RHIC, is observed. 

{\bf 5.} Summary: In this paper we make predictions for soft hadronic observables that will be soon measured in Pb+Pb collisions at $\sqrt{s_{\rm NN}}$ = 2.76 TeV. Our results are based on the hydrodynamic model, which was 
successfully used  to describe the RHIC results. Compared to the RHIC data, the values of various physical parameters show smooth changes with the beam energy, however, our predictions indicate larger changes of the multiplicity than those expected earlier. The main reason for this fact is the large multiplicity observed in $pp$ collisions at the LHC, which is used as the input for our estimate.  Another input value that favors the increase of the multiplicity is a relatively large value of the parameter $\alpha$. This leads to larger values of the mean transverse momentum of emitted particles and of the integrated elliptic flow.

We hope that our results will be useful for future comparisons between theory and experiments at the LHC. 

{\bf Acknowledgments: }  We thank Adam Kisiel for discussions which led
 us to the recognition of the problems discussed in this paper. Research 
supported in parts by the MNiSW grant No. N N202 263438, the Foundation
 for Polish Science, and Slovak-Polish Collaboration grant No.~SK-PL-0021-09.



\end{document}